\documentclass[aps,superscriptaddress,showpacs]{revtex4}
\usepackage{amssymb}
\usepackage{amsmath}
\usepackage{amsthm}
\newcommand\scri{\mathcal{I}}

\newtheorem{thm}{Theorem}
\newtheorem*{cor}{Corollary}
\newtheorem*{corTC}{Corollary (Topological Censorship)}

\theoremstyle{remark}
\newtheorem*{rem}{Remark}

\begin{document}

\title{Topological Rigidity of Dimensional Reduction to Three Dimensions}
\author{J.\ Brannlund}
\author{S.\ Slobodov}
\author{K.\ Schleich}
\author{D.M.\ Witt}
\affiliation{Department of Physics and Astronomy, University of British Columbia,
Vancouver, British Columbia \ V6T 1Z1}
\date{\today}

\begin{abstract}
Studying spacetimes with continuous symmetries by dimensional reduction to a lower dimensional spacetime is a well known technique in field theory and gravity.  Recently, its use has been advocated in numerical relativity as an efficient computational technique for the numerical study of axisymmetric asymptotically flat 4-dimensional spacetimes. We prove here that if the dimensionally reduced spacetime is a physically reasonable 3-dimensional asymptotically flat or asymptotically anti-de Sitter spacetime, then, surprisingly, the topology of the higher dimensional spacetime must be one of two product topologies. Reductions of other topologies result in physically pathological spacetimes. In particular, reduction of asymptotically flat 4-dimensional spacetimes must lead to pathologies. These results use only the topological censorship theorem and topological
methods and consequently are independent of the field equations and reduction method.  

\end{abstract}
\maketitle
Many commonly studied spacetimes have a continuous global symmetry group  $G$. In certain contexts, it is useful to study these spacetimes through a reduced spacetime $M/G$ where points of the spacetime manifold $M$ related 
by the group action are identified and $G$ invariant tensors on $M$ projected into $M/G$ yield the reduced spacetime's geometry and effective matter sources. 
Such dimensional reductions have long been utilized in Kaluza Klein models to relate the physics of higher dimensional backgrounds to that of 4-dimensional spacetime. In classical gravity, reduced spacetimes have also been used as a tool for the generation of new 4-dimensional solutions \cite{ehlers,Geroch:1970nt,Moncrief:1986}. More recently in numerical relativity,  dimensional reduction has been used to reduce the computational complexity of the Einstein equations  for axially symmetric 4-dimensional spacetimes such as axisymmetric collapse \cite{Choptuik:2003ac}.  This approach, termed the  $2+1+1$ formulation, reduces the Einstein equations to those for a 3-dimensional spacetime by removing a 
$U(1)$ killing vector and inducing equivalent matter sources. These reduced Einstein equations are then cast into the initial value formulation  and initial data for this 3-dimensional spacetime is then numerically evolved. The axially symmetric 4-dimensional spacetime is then constructed from this evolution. 
%%This technique has been advocated as an efficient method of numerically studying important astrophysical problems with axial symmetry such as black hole collisions and axisymmetric collapse.

Given the broad application of dimensional reduction, it is natural to ask what relationships exist between properties of the original and reduced spacetimes. It is well known that dimensionally reduced spacetimes can have mathematical and physical characteristics that are different than those of the unreduced spacetime. For example, dimensional reduction of nonsingular spacetimes can result in singularities  \cite{Sorkin:1983ns,Gross:1983hb,Gibbons:1994vm} and certain Kaluza-Klein solutions are known to correspond to zero or negative mass 4-dimensional spacetimes \cite{Witten:1981gj,Brill:1989di}.  However, little systematic examination of the relation of mathematical and physical properties of the original spacetime to those of the reduced spacetime has been carried out. In particular,  the Cauchy problem for general relativity  \cite{he2} for the reduced spacetime is especially relevant to the $2+1+1$ formulation
as the reduced spacetime is constructed by numerical evolution from initial data. The Cauchy problem for general relativity, that is the existence, uniqueness  and continuous dependence on initial data of a solution to the initial value problem,  requires certain conditions,  in particular suitably regular and differentiable initial data and the dominant energy condition (DEC). 
 However, not all dimensional reductions of  a spacetime result in a  reduced spacetime that satisfies these conditions.
 
For example, consider the reduction of the Reissner-Nordstrom metric
$$ds^2 = - V(r) dt^2 + \frac {dr^2}{V(r)} + r^2 (d\theta^2 + \sin^2 \theta d\phi^2),$$
$V(r) = (1-\frac {2M}r + \frac {Q^2}{r^2})$
by the $U(1)$ symmetry generated by  $\xi^a=(\partial/\partial \phi)^a$.  Reduction by the projection $h_{ab} = g_{ab}-\frac 1\lambda \xi_a\xi_b$ where  $\xi^a\xi^b g_{ab} = \lambda=r^2\sin^2\theta$ yields an asymptotically flat 3-dimensional spacetime
$$ ds^2 = - V(r) dt^2 + \frac {dr^2}{V(r)} + r^2 d\theta^2 $$
It violates all energy conditions outside the black hole horizon for $Q<M$,  $M$ positive, as 
all observers see negative energy density;
$$u^au^bT_{ab} = u^au_a (\frac M{r^3} - \frac {Q^2}{r^4}) + (u^\theta)^2 (-\frac {3M}{r^3}+ \frac {4Q^2} {r^4})
$$ where  $u^au_a = -1$, $u^\theta$ the  $\theta$-component of $u^a$ in the orthonormal frame. This violation extends arbitrarily far into the asymptotic region. 
%%  (In contrast, note that all energy conditions are satisfied everywhere if $M$ is negative; of course, in this case the spacetime exhibits a naked singularity at $r=0$.)
 Thus, as DEC is violated, this reduction results in a 3-dimensional spacetime for which the Cauchy problem does not apply to its formulation as a $2+1$ initial value problem.

Of course, different projections  yield  different reductions. In particular, it is well known that  judicious conformal rescaling of the metric, $\bar h_{ab}= \lambda h_{ab}$, leads to 3-dimensional Einstein equations for $\bar h_{ab}$ with effective matter sources that satisfy DEC. However, a direct calculation of the scalar curvature of the rescaled metric yields a naked timelike singularity at $\theta = 0$ and $\pi$, even for $Q=M=0$, the case of a reduced Minkowski spacetime. Thus this reduction also yields a 3-dimensional spacetime for which the Cauchy problem fails to be well posed;  any evolution of such data must involve solving in the presence of naked singularities. 
In addition, the rescaled metric is not asymptotically flat for this reduction, making it problematic to define asymptotically defined quantities such as mass and charge.

Clearly, given that certain reductions lead to lower dimensional spacetimes with pathologies, it is very interesting to determine under what conditions a reduction exists that lead to no pathologies.  In this paper we answer this question for all reductions that lead to a 3-dimensional asymptotically flat or asymptotically anti-de Sitter spacetime. We prove, surprisingly, that  a topological restriction exists;  for the reduced spacetime to be physically reasonable, i.e.  globally hyperbolic and satisfying the null energy condition,  the original spacetime must have product topology of  ${\mathbb R}^3\times G$ or  ${\mathbb R}^2\times S^1\times G$ where $G$ is the connected symmetry group. Hence any reduction of a higher dimensional spacetime with $G$ not of these product topologies must result in pathologies such as those illustrated in  the Reissner-Nordstrom case. 
Our theorem follows as a consequence of the topological censorship
theorems \cite{Friedman:1993ty,Galloway:1999bp,Galloway:1999br} and techniques from algebraic 
topology.  The theorem does not utilize details of either the reduction or the field equations in either the unreduced or reduced spacetime. Hence the results are generic.

We begin with necessary definitions:
Given a spacetime $M$ with a group  of spacelike symmetries $G$ generated by Killing 
vectors, a {\it reduction} will be the lower dimensional spacetime $M/G$ 
with dimension $dim(M/G)=dim(M)-dim(G)$. A {\it bundle reduction} is the case in which $M/G$ is the 
base space of $G$-bundle with total space $M$. An example of a bundle reduction is that utilized in \cite{Geroch:1970nt}  where $M$ is 4-dimensional spacetime and $G=U(1)$ or $\mathbb R$. 
A {\it singular bundle reduction} is the case in which $M$ fails to be a fiber bundle at certain points. 

A spacetime satisfies the {\it null energy condition} (NEC) if $R_{ab} W^a W^b \geq 0$ 
for all null $W^a$, the {\it weak energy condition} (WEC) if $T_{ab} W^a W^b \geq 0$ 
for all timelike $W^a$  and the {\it dominant energy condition} (DEC) if WEC holds and
$T_{ab} W^b T^a_{c} W^c \leq 0 $ \cite{he}.

A spacetime $( M, g_{ab})$ is {\it asymptotically flat} (AF) if it can be
conformally included into a spacetime-with-boundary
$M'=  M \cup \scri$, with metric $g_{ab}'$, such that (a)
 for some conformal
factor $\Omega\in C^{1}(\cal M')$, $g_{ab}' = \Omega^2 g_{ab}$ on $M$  and $\Omega$ vanishes on $\scri$ but has null gradient which is nonvanishing  pointwise on $\scri$. (b) The boundary $\partial 
M' = M'\setminus M= \scri$ is a disjoint union of past and future parts $\scri^+ \cup\scri^-$, each having topology $S^2\times \mathbb R$ with $\mathbb R$'s complete null generators. 
A spacetime $(M, g_{ab})$ is {\it asymptotically locally anti-de\! Sitter} (ALADS) if it can be
conformally included into a spacetime-with-boundary
$M'=  M \cup \scri$, with metric $g_{ab}'$, such that $\partial 
M' = \scri$ is timelike ({\it i.e.}, is a Lorentzian hypersurface 
in the induced metric) and $M = M' \setminus \scri$. 
The conformal
factor $\Omega\in C^{1}(\cal M')$ satisfies (a) $\Omega > 0$ and $g_{ab}' = \Omega^2 g_{ab}$  on $M$, and (b) $\Omega = 0$ and $d \Omega \ne 0$ pointwise on $\scri$. We permit $\scri$ to have multiple components. 

 A spacetime $M'$ with boundary is {\it globally hyperbolic} if it is strongly causal and the sets $J^+(p,{\cal M'})\cap 
J^-(q,\cal M')$ are compact for all $p,q\in \cal M'$ \footnote{The timelike future (causal future)
of a set $S$ relative to  $U$, $I^+(S,U)$ ($J^+(S,U)$), is the 
set of all points that can be reached from $S$ by a future directed  
timelike curve (causal curve) in $U$.  The interchange of the past with future in the previous
definition yields $I^-(S,U)$ ($J^-(S,U)$). }. A   {\it Cauchy surface} $V'$
is a spacelike hypersurface 
such that every non-spacelike curve intersects this surface exactly once. A {\it partial Cauchy surface} is a surface that satisfies the weaker condition that each non-spacelike curve intersects 
the surface at most once. Note $V'$ for a manifold with boundary will have boundary on $\scri$ \footnote{These definitions are the usual extensions of those for manifolds without boundary used in proofs of topological censorship for locally asymptotically anti-de Sitter spacetimes.}.
The {\it domain of outer communications} (DOC) is the portion of a spacetime  ${ M}$ which is exterior to event horizons. Precisely ${D} = I^-(\scri^+_0)\cap I^+(\scri^-_0)$ for a connected component $\scri_0$ for an AF spacetime and
${D} = I^-(\scri_0)\cap I^+(\scri_0)$ for an ALADS spacetime is the subset of ${ M}$ that
is in causal contact with $\scri$ 
Note that $D$ is the interior of an $(n+1)$-dimensional spacetime-with-boundary
 ${D}' = {D }\cup \scri$ and that $D'$ is itself a globally hyperbolic spacetime with boundary.

The first theorem is our main result: %for bundle reductions:

\begin{thm} Let $D^{2+1+k}$  be the DOC of a globally hyperbolic spacetime with boundary with 
a spacelike connected symmetry group, $G$. If $D^{2+1+k}$ has a globally hyperbolic bundle
reduction  $\Sigma^{2+1}={{D^{2+1+k}}/ {G}}$ satisfying both of the following conditions:

i) $\Sigma^{2+1}$ is AF or ALADS.

ii) The induced metric on $\Sigma^{2+1}$ satisfies NEC.

\noindent Then  ${D^{2+1+k}}= {\mathbb R}^3\times G$ or 
${D^{2+1+k}}= {\mathbb R}^2\times S^1\times G$.
\end{thm} 

To prove this theorem  we will utilize a key corollary of the topological censorship theorem  for 3-dimensional spacetimes \cite{Galloway:1999br}. 

\begin{corTC}  Let $D^{2+1}$ be the DOC of a globally hyperbolic AF or ALADS spacetime with boundary satisfying NEC, 
and let $V$ be a Cauchy surface in $D^{2+1}$. Then  $V$ is either $B^2$ (a disk) 
or $I\times S^1$ (an annulus).
\end{corTC} 

\begin{proof} (We provide here an alternate proof here to that in \cite{Galloway:1999br}.)  Let $V'$ be the 2-dimensional Cauchy surface in Penrose compactified
spacetime $D'=D\cup\scri$. with boundary at spatial infinity $\partial V =\Sigma_{\infty}$, as described above. As $D'$ satisfies the conditions needed to prove topological censorship, it follows that the homomorphism of fundamental groups $i_*: \Pi_1(\Sigma_{\infty}) \to \Pi_1(V')$ induced by inclusion is surjective \cite{Galloway:1999bp,Galloway:1999br}.

Now, in 2 dimensions, the only choice for the spatial boundary topology at infinity is $S^1$, the only closed connected 1-dimensional manifold. Thus $\Pi_1(S^1) \to \Pi_1(V')\to 1$. 
Since $\Pi_1(S^1) ={\mathbb Z}$, it follows that  ${\mathbb Z} \to \Pi_1(V')\to 1$. This exact 
sequence  and basic group theory imply that  $\Pi_1(V')= {\mathbb Z}/{\rm ker\  i_*}$. Since the 
kernel of $i_*$ must be a subgroup of the integers, 
${\rm ker\  i_*}\subset {\mathbb Z}$,  it follows that ${\rm ker\  i_*} = s {\mathbb Z}$ where s is a fixed non-negative 
integer. Therefore, $\Pi_1(V')= {\mathbb Z}/ s {\mathbb Z}= {\mathbb Z}_s$. Now, from the classification of 2-manifolds, $V$ must be a closed orientable 2-manifold  minus a disjoint
union of one or more disks.  It follows from the classification of 2-manifolds and their fundamental groups that $s=0$ or $s=1$ and
that the only choices for $V$ are $B^2$ (a disk) 
or $I\times S^1$ (an annulus).
\end{proof}
\begin{rem} One can also consider the more general case of non-compact boundary, $\Sigma_{\infty}={\mathbb R}$. As topological censorship holds under NEC without assumption of compactness of $\Sigma_{\infty}$, it again follows that $i_*: \Pi_1({\mathbb R}) \to \Pi_1(V')\to 1$ which, in this case, yields $\Pi_1(V') =1$. Hence $V$ is contractible, hence is ${\mathbb R}^2$.
\end{rem}

The proof of Theorem 1 will utilize the universal classifying space $BG$ of the Lie group $G$. 
Given the universal classifying space $BG$, there is a one to one correspondence between
$G$-bundles over a space $X$ and homotopy classes of maps $[X, BG]$ from $X$ to
$BG$. This correspondence is related to classification of bundles by characteristic classes such as the Chern class, however, the construction using the classifying space $BG$
is more general. 

The universal classifying space $BG$ of a Lie group $G$ is the base space of a principal $G$-bundle
with contractible total space $EG$. For any Lie group $G$ there exists $BG$  \cite{hues}. 
Typically, the universal classifying space is infinite dimensional \footnote{However note that the Milnor 
construction of $BG$ does not yield an infinite dimensional manifold.}. The universal classifying space $BG$ is 
unique up to homotopy type. This fact enables one to utilize, if convenient,  a homotopy equivalent space which has a Hilbert manifold structure \footnote{An example of a universal classifying space relevant to $U(1)$ bundle reductions is $BU(1)$. A homotopy equivalent space with Hilbert manifold structure is ${\mathbb C}P^{\infty }$. An explicit construction of this space can be obtained by letting $U(1)$ act on the Hilbert space ${\it l}^2({\mathbb C})$ (a contractible space) by multiplication by a phase.  Identification of any two non-zero vectors which differ by a phase yields ${\mathbb C}P^{\infty }$, Clearly $BU(1)={\mathbb C}P^{\infty }$. One can similarly construct  $BSU(2)={\mathbb H}P^{\infty }$. }.

\begin{proof}[Proof of Theorem 1]
As $G$ is spatial, ${D^{2+1+k}}/ {G}$ is the DOC of a 3-dimensional spacetime that is AF or ALADS by conditions of the theorem. 
The Topological Censorship corollary applied to ${D^{2+1+k}}/ {G}$ directly implies that
${D^{2+1+k}}/ {G}={\mathbb R}^3$ or ${D^{2+1+k}}/ {G}={\mathbb R}^2\times S^1$.
If ${D^{2+1+k}}/ {G}={\mathbb R}^3$, then the $G$-bundles over ${D^{2+1+k}}/ {G}$ are given by $[{\mathbb R}^3,BG]=[\star, BG]=1$ where $\star$ is a fiducial point
because ${\mathbb R}^3\simeq \star$. Hence, there is only one $G$-bundle, the trivial product bundle.
Therefore, ${D^{2+1+k}}={\mathbb R}^3\times G$ for this case. 

Next, for ${D^{2+1+k}}/ {G}={\mathbb R}^2\times S^1$ the $G$-bundles are again
given by $[{\mathbb R}^2\times S^1,BG]=[S^1, BG]=\pi _1(BG)$ because  
${\mathbb R}^2\times S^1\simeq S^1$. To calculate, $\pi _1(BG)$ we use the following exact
sequence for  fibrations  \cite{white} for
%%$$\dots \rightarrow \pi _k(F) \rightarrow \pi _k(E)\rightarrow \pi _k(B)\rightarrow 
%%\pi _{k-1}(F)\rightarrow $$
%%$$\dots \rightarrow \pi _0(F) \rightarrow \pi _0(E)\rightarrow \pi _0(B)\ .\label{exseq}$$
 the G-bundle  $p:EG\rightarrow BG$ with fiber $G$, 
%%so (\ref{exseq}) explicitly for our case yields,
$$\dots \rightarrow \pi _k(G) \rightarrow \pi _k(EG)\rightarrow \pi _k(BG)\rightarrow 
\pi _{k-1}(G)\rightarrow $$
$$\dots \rightarrow \pi _0(G) \rightarrow \pi _0(EG)\rightarrow \pi _0(BG)\ .$$ As $EG\simeq \star$, the sequence yields
$1\rightarrow  \pi _1(BG)\rightarrow \pi _0(G)\rightarrow 1.$ Thus $\pi _1(BG)=\pi _0(G)$. Now $G$
is connected so  $\pi _0(G)=1$. Hence, $\pi _1(BG)=1$. Therefore, again there is only the trivial product bundle, ${D^{2+1+k}}={\mathbb R}^2\times S^1\times G$.
\end{proof}

Note that Theorem 1 depends critically on the dimension. Topological censorship holds in 3 or more spacetime dimensions, not in 2. Hence, evolution problems reducible to 2 spacetime dimensions, i.e. those with spherical symmetry, have no topological restriction from the above techniques. 

The Topological Censorship corollary shows that 3-dimensional gravity is topologically rigid; there can be at most one black hole horizon in each asymptotic region of a 3-dimensional spacetime. 
Thus not only are geons hidden behind horizons in 
3-dimensional gravity, there are also no AF or ALADS multiple black hole solutions. This topological rigidity is key to the proof of Theorem 1 and the topological rigidity of dimensional reduction.

A key fact is that Theorem 1 does not require any particular energy condition or asymptotic condition on the higher dimensional spacetime. However, further results follow if such a condition is imposed, namely
\begin{cor} There is no  $U(1)$ bundle reduction of a 4-dimensional AF globally hyperbolic spacetime to a 3-dimensional AF or ALADS spacetime that satisfies NEC. 
\end{cor}
If $M^{2+1+k}$ is a globally hyperbolic spacetime satisfiying NEC, topological censorship implies that its DOC is simply connected. In 4 dimensions,  Theorem 1 implies $D^{4}= {\mathbb R}^3 \times G$ or $S^1\times {\mathbb R}^2 \times G$, neither of which are simply connected for $G= U(1)$ as $G$ itself is not simply connected.\qed

Note that if $G = \mathbb R$, then $D^{4}=  S^1\times {\mathbb R}^2 \times G$ is the DOC of the  4-dimensional black string solution. Of course, this spacetime is not AF and is thus not a counterexample to the corollary.

In addition, no such corollary holds for the ALADS case; the 4-dimensional AdS soliton \cite{Horowitz:1998ha}
$$ds^2 = - \frac{ r^2}{l^2} dt^2 + \frac {dr^2}{V(r)} +\frac{ r^2}{l^2} d\theta^2 + V(r)d\phi^2,$$
$V(r)= \frac {r^2}{l^2}(1- \frac {r_0^3}{r^3})$
has a reduction satisfying Theorem 1. The reduction $h_{ab} = g_{ab}-\frac 1\lambda \xi_a\xi_b$  with  $ \xi_a=(\partial/\partial \theta)^a$ results in an asymptotically AdS spacetime
$$ds^2 = - \frac{ r^2}{l^2} dt^2 + \frac {dr^2}{V(r)}  + V(r)d\phi^2$$ that yields
$$u^au^bT_{ab} = u^au_a (\frac 1{l^2} + \frac 12\frac {r_0^3}{l^2r^3}) +\frac 32 (u^t)^2 \frac {r_0^3}{l^2r^3}$$ which is positive for null $u^a$, $u^au_a = 0$.
Hence, the spacetime satisfies NEC everywhere. The reduced spacetime is a bundle reduction with topology $B^2 \times {\mathbb R}$. 

Although the original reason for the $2+1+1$ formalism in numerical relativity was to simplify numerical calculations, the work presented here shows reduction of an AF spacetime invariably induces a pathological 3-dimensional spacetime. Clearly, Theorem 1 and its corollary show that Cauchy problem does not apply in  the $2+1+1$ formalism. Special handling may yet allow application of this formalism; for instance, one can utilize reduction techniques that yield a naked singularity if one appropriately controls it in some fashion, for example by requiring that it be regular in the reconstructed 4-dimensional spacetime. However, the mathematical rigor of such solution techniques remains to be proven.  Furthermore numerical simulations of  axisymmetric black hole collisions by reduction might be particularly problematic as they must have 2 or more black hole horizons, a direct violation of topological rigidity. In contrast to AF spacetimes, reductions of 4-dimensional ALADS spacetimes   are allowed as they can have more general topology as illustrated by the AdS soliton. Thus a small negative $\Lambda $ could be introduced  to ``regulate"
pathologies in reductions of asymptotically vacuum spacetimes and the limit $\Lambda \rightarrow 0$ taken. However, such an approach is still subject to the requirement that the spacetime has an allowed topology. As seen above, both of these topologies are incompatible with the topology of $\scri $ for 4-dimensional AF spacetime, so such a regulation cannot fix this particular problem.

When the reduced spacetime is of 4 or more dimensions, topological censorship will again restrict the topology of the higher dimensional spacetime.  In these cases however,  there is much more freedom in allowed 4-dimensional spacetime topologies. In 4-dimensional AF spacetimes, the DOC is restricted to ${\mathbb R}^3$ minus arbitrary numbers of 3-balls. ALADS 4-dimensional spacetimes  allow even more general topology as  $\scri$ can have the topology of any 3-manifold. Hence, the analogues of Theorem 1 need be proven for cases, rather than in general.

Finally, Theorem 1 assumes that the reduction is a bundle. This may not  be the case if the Killing
vectors generating the global symmetries have zeros. Zeros of the Killing vectors correspond to points in $M$ being fixed under the group action $G$. At these points, the reduction may fail to be a fiber bundle. However, the proof of Theorem 1 can be extended to cover such cases. The details of this  proof  will be presented elsewhere \cite{future}, but  below we will outline the proof of the special case of singular fiber bundles, Siefert fiber manifolds, with fiber $G=U(1)$.  

For this case, the original spacetime is 4-dimensional and its Cauchy surface is a 3-manifold foliated by closed fibers $U(1)$. This fibered space can either be a fiber bundle (in which case Theorem 1 applies) or a singular fiber bundle also called a {\it Seifert fiber manifold} or {\it Seifert fiber manifold}  \cite{jaco} . It differs from a non-singular bundle by the
following local effect: each fiber in the singular fiber bundle has a neighborhood diffeomorphic to one in a model Seifert fibering of $S^1 \times B^2$. A model Seifert fibering of $S^1 \times B^2$ is constructed from  $[0,1] \times D^2$ by identifying the disks
$\{ 0 \} \times D^2$ and  $\{ 1\} \times D^2$  via a $2\pi p/q$ rotation, $p, q$ relatively prime integers. The segment $[0,1] \times {0}$ becomes the fiber $S^1 \times {0}$, while every other fiber is made from $q$ segments $[0,1]$. Intuitively, the fiber at the origin is fixed and the fibers surrounding it are rotated. Due to this rotation, a neighborhood of the fiber at the origin is not diffeomorphic to a neighborhood in a principle bundle of $U(1)$ over $B^2$ and hence this model Seifert fibering is not a fiber bundle. A Seifert fiber bundle can have more than one such singular fiber.  

For Seifert fiber bundles, the Topological Censorship corollary again implies
that the spatial topology of the reduced DOC is a disk or annulus. One can next use theorems from 3-manifold
topology to conclude that every Seifert fiber manifold with boundary is irreducible; that is every 2-sphere
bounds a 3-ball. This implies that $\pi _2(D^{2+1+1})=0$. Therefore, it follows that $D^{2+1+1}={\mathbb R}^3\times U(1)$. These results can also be generalized to the case of general $G$ and to the case of ALADS boundary conditions; however, the topological arguments are more complex than those outlined  above.

The authors KS and DW would like to thank NSERC for its support and the Perimeter Institute for Theoretical Physics for its hospitality during the initiation of this work. 
SS would like to thank NSERC for PGS support.

\end{document}